\documentclass{vgtc}                          

\graphicspath{{figures/}{pictures/}{images/}{./}} 

\usepackage{times}                     

\usepackage{tabu}                      
\usepackage{booktabs}                  
\usepackage{lipsum}                    
\usepackage{mwe}                       

\usepackage{mathptmx}                  
\usepackage{amsmath}
\usepackage{amsfonts}
\usepackage{xcolor}
\usepackage{enumitem}

\newcommand\blfootnote[1]{%
  \begingroup
  \renewcommand\thefootnote{}\footnote{#1}%
  \addtocounter{footnote}{-1}%
  \endgroup
}
\def\rev#1{{\color{black} #1}}

\onlineid{0}

\vgtccategory{Research}

\vgtcinsertpkg

\title{Distribution-Agnostic Isocontour Confidence Bounds for Robust Uncertainty Visualization of Scalar Field Data}

\author{Timbwaoga A. J. Ouermi \thanks{e-mail: ouermita@ornl.gov}\\
        \scriptsize \centering Oak Ridge National Laboratory %
\and Nina M. Gottschling \\ 
     \scriptsize \centering  Oak Ridge National Laboratory %
\and Alex Gorczowski \\ 
     \scriptsize \centering U. of I. Urbana-Champaign
\and Tushar M. Athawale \\ 
     \scriptsize \centering Oak Ridge National Laboratory
     }

\teaser{
  \centering
  \includegraphics[width=0.90\linewidth]{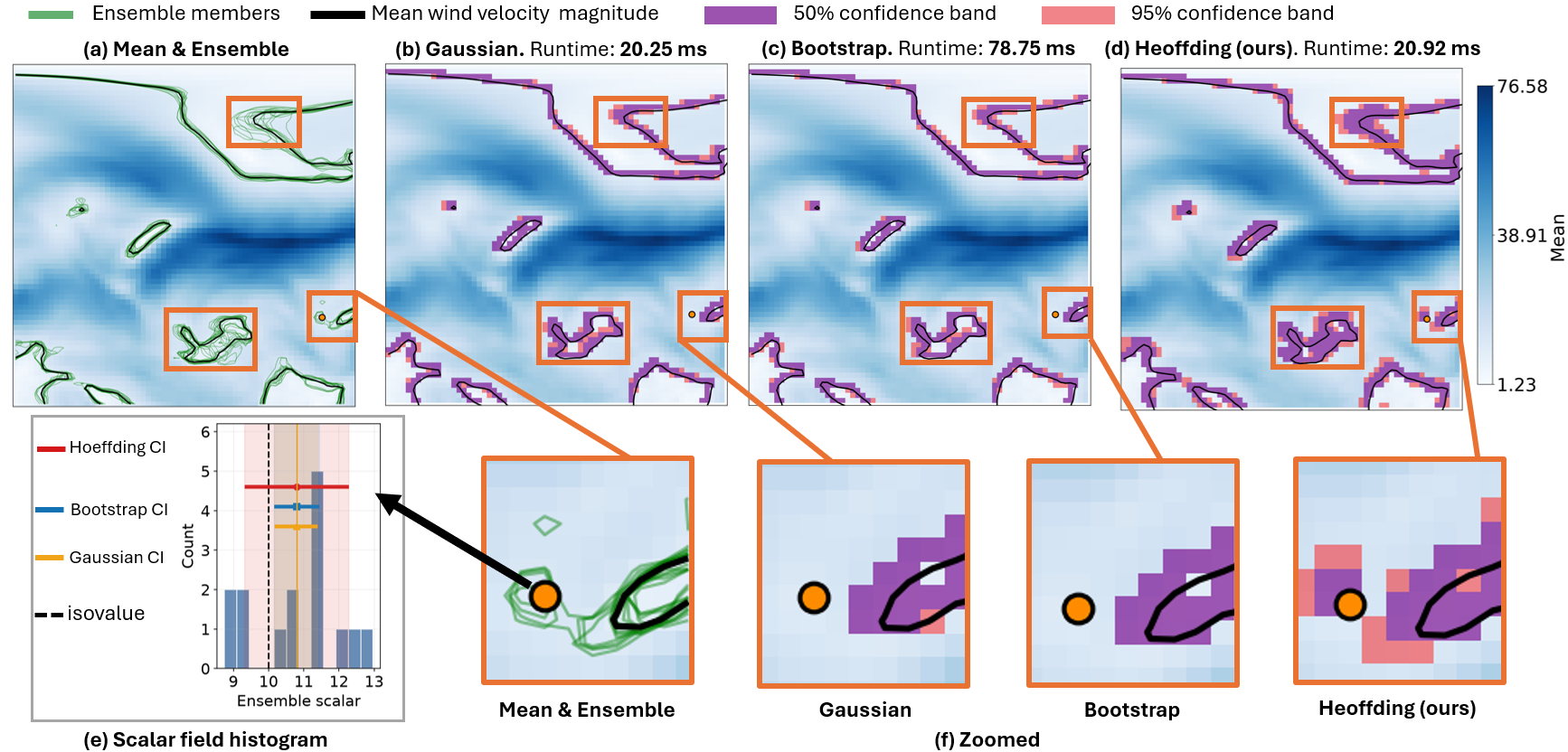}
  \vspace{-3mm}
  \caption{
  Robust distribution-agnostic Hoeffding confidence band visualization for uncertain scalar field. The zoomed-in regions clearly highlight enhanced confidence band depiction using our distribution-agnostic approach based on Hoeffding's inequality (d), compared to traditional distribution-driven (b) and nonparametric bootstrap (c) methods, with the ensemble contours visualized in green (a) as the reference. In all images, the mean vector field magnitude is colormapped in blue with the mean isocontour depicted in black. Gaussian and bootstrap models can lead to narrowing of contour-crossing bands (see zoomed views) due to their poor representation from the limited number of samples, as seen in the histogram (e). In contrast, the Hoeffding confidence band in (d) is wider and provides a robust, distribution-agnostic confidence band. The proposed Hoeffding method is efficient and has comparable runtime ($20.92$ ms) with respect to the Gaussian method and substantially lower than that of bootstrap model.
  }
  \label{fig:teaser}
}

\abstract{

Uncertainty visualization has been shown to be pivotal for conveying the reliability of features extracted from scalar fields. Features represented by individual isocontours and mean isocontours lack an indication of spatial uncertainty, whereas spaghetti isocontour plots can become cluttered and difficult to interpret. Existing methods relying on specific distribution assumptions, such as Gaussian and nonparametric bootstrap, provide compact, clutter-free spatial confidence bounds but may underestimate uncertainty for ensembles with a limited number of samples. We introduce a robust, distribution-agnostic {\em Hoeffding confidence band} as a novel complementary (and not competitive) technique to mitigate potentially misleading uncertainty bounds that may arise from distribution-based assumptions. The approach constructs vertex-wise confidence bounds using Hoeffding's inequality and propagates them to generate isocontour confidence bands. Results on synthetic and real ensemble datasets show that the Hoeffding confidence bands are loose but accurately capture underlying true values that may be missed by the Gaussian and bootstrap alternatives, while remaining computationally efficient.
} 


\begin{document}



\maketitle
%
\section{Introduction} 

Scalar field visualization often drives interpretation and decision-making by revealing level-set structures such as fronts, interfaces, and coherent features; however, due to inherent uncertainty in the underlying data (e.g., uncertainty in simulation output from the choice of initial and boundary conditions, spatial discretization, and data quantization), these extracted structures are often uncertain as well. \blfootnote{This manuscript has been co-authored by UT-Battelle, LLC, under contract DE-AC05-00OR22725 with the US Department of Energy (DOE). The US government retains and the publisher, by accepting the article for publication, acknowledges that the US government retains a nonexclusive, paid-up, irrevocable, worldwide license to publish or reproduce the published form of this manuscript, or allow others to do so, for US government purposes. DOE will provide public access to these results of federally sponsored research in accordance with the DOE Public Access Plan (http://energy.gov/downloads/doe-public-access-plan).}Common strategies to communicate ensemble data include visualizing the mean, ensemble, or probabilistic level-set representations. The mean contours provide a compact deterministic summary, but they do not communicate confidence about the isocontour positions, potentially leading to data misinterpretation. Overlaying all ensemble contours (also known as a spaghetti plot)~\cite{TA:Potter:2009:ensemble-vis} reveals their spatial variability, but they can become cluttered and therefore difficult to interpret. Prior works have addressed these limitations by assuming diverse probabilistic distributions, such as Gaussian~\cite{Athawale2026,Pothkow2011,Wang2023}, uniform~\cite{Athawale2013}, and nonparametric~\cite{Athawale2016,athawale2025,Pothkow2013}, to characterize and visualize uncertain scalar values.

One of the major limitations of distributional assumptions or empirical/nonparametric approaches in state-of-the-art techniques is that {\em distributional forms of ensembles are often unknown, and in many cases not reliably captured in practical scenarios where ensemble sizes are typically small ($n \le 100$)}. In other words, a small sample size often poorly captures the true shape of the probability distribution. Under these conditions, we propose a novel utilization of {\em distribution-agnostic} concentration inequalities as a principled alternative to existing distribution-driven and empirical approaches. From previous literature, Hoeffding’s inequality \cite{hoeffding1994probability} offers bounds for sample means of bounded independent and identically distributed (i.i.d.) random variables without requiring knowledge of the underlying distribution, while McDiarmid’s inequality \cite{mcdiarmid1989method} extends similar guarantees to functions of i.i.d. variables. When variance information is available, Chebyshev’s inequality \cite{tchebichef1874valeurs} provides an additional alternative. In this work, we leverage Hoeffding’s inequality~\cite{hoeffding1994probability} for deriving robust isocontour confidence bands.



A key advantage of these distribution-agnostic inequalities is that they enable {\em efficient estimation of theoretical uncertainty bounds} directly from sparse samples, without knowledge of the underlying distribution. Although they can yield loose error bounds, they {\em guarantee to encompass unknown truth in the computed bounds}. This makes these techniques a robust and trustworthy baseline or safety-oriented visualization method, particularly when the number of ensemble members is limited in real applications such as expensive fluid dynamics simulations. In summary, our contributions are threefold: (1) We derive data and isocontour confidence intervals representing their uncertainty through a novel application of distribution-agnostic Hoeffding's inequality; (2) We showcase how our distribution-agnostic approach provides, although loose, reliable confidence bounds encompassing unknown truth not captured by distribution-driven and bootstrap approaches, especially in realistic scenarios where the number of ensembles is limited; (3) We demonstrate the effectiveness of our methods through experiments on synthetic and real wind flow and K\'{a}rm\'{a}n vortex datasets.




%
\section{Related Work}
\rev{There have been significant advances over the past decade to quantify and visualize uncertainty in scalar field data~\cite{TA:Bonneau:2014:StateOftheArtUQ,pang1997approaches}. For level-sets, probabilistic marching cubes (MC) estimates uncertain contour crossings~\cite{Pothkow2011}, while related approaches study MC cases uncertainty~\cite{marchingcubesTopoUncertainty}, isosurface variability~\cite{Pfaffelmoser2011}, interpolation uncertainty~\cite{Athawale2013}, nonparametric uncertainty~\cite{Pothkow2013}, spatially correlated Gaussian uncertainty~\cite{Athawale2026}, and AI-/information-driven algorithms~\cite{Han2022, data-driven-pmc}. In addition, uncertainty techniques for direct volume rendering~\cite{Djurcilov2001}, probabilistic volume animation~\cite{Lundstrom2007}, statistical volume rendering~\cite{Sakhaee2017}, and nonparametric volume rendering~\cite{Athawale:2021:AJE} have been developed. Feature-based methods address uncertain critical points ~\cite{athawale2025,Liedmann2016}, vector-field features~\cite{Petz2012}, Morse complexes~\cite{TA:2022:Athawale:UQMorseComplex}, and merge trees~\cite{mergeTreeUncertainty}. These methods fit specific distribution models to ensemble data to compute and visualize uncertainty.}


%

A few empirical methods have been developed, including contour boxplots~\cite{Whitaker2013}, probabilistic inclusion depth~\cite{Chaves-de-Plaza2024,probabilisticInclusionContour}, curve boxplots~\cite{Mirzargar2014curve}, \rev{and curve density estimates~\cite{Lampe2011}},  which derive contour depth rank statistics directly from ensemble members to quantify quantiles and confidence levels, but do not provide theoretical guarantees on unknown true data or contours. 
In contrast, concentration-inequality-based bounds such as Hoeffding's~\cite{hoeffding1994probability} take a different approach in which {\em they provide theoretical guarantees on unknown truth for the ensemble}. Specifically, rather than deriving confidence bounds probabilistically or empirically, Hoeffding's inequality yields theoretical uncertainty bounds that encompass the unknown true data. Our work leverages this direction, applying Hoeffding's inequality to construct a distribution-agnostic, theoretical confidence bands as a novel complementary (and not competitive) method for stable analysis of uncertain isocontours.


%
\section{Hoeffding Confidence Band}
\label{sec:hoeffding-uncertainty-bound}
We now present our method, which leverages Hoeffding's inequality to construct uncertainty bounds for scalar data and derive isocontour uncertainty, referred to as \textit{Hoeffding confidence band}. Given independent random variables $X_1$, $X_2$, $\dots$, $X_n$ where each observation is bounded $X_i \in [a,b]$. Hoeffding's inequality gives a theoretical, distribution-agnostic bound on deviation $\epsilon$ of the empirical mean \(\hat{\mu}_n =  \frac{1}{n}\sum_{i=1}^{n} X_i\) from the (usually unknown) ground truth \(\mu = \mathbb{E}[X_1]\) \rev{of the generating distribution}, expressed as follows:
\begin{equation}\label{eq:hoeffding-ineq}
\Pr\left( \left|\hat{\mu}_n - \mu\right| \ge \epsilon \right)
\le
2\exp\left(-\frac{2n\epsilon^2}{(b-a)^2}\right).
\end{equation}
\rev{The inequality in \cref{eq:hoeffding-ineq} is powerful in that it provides a pointwise confidence bound regardless of the underlying probability distribution of the data. The only requirement is that the samples are independent and bounded ($X \in [a,b]$).}

In our derivation, the Hoeffding confidence band for an isocontour for a given confidence level $\beta$ is constructed in two stages. The first stage consists of computing the confidence interval for the scalar mean value at each grid vertex using Hoeffding's inequality in \cref{eq:hoeffding-ineq}. The second stage consists of calculating the cell-crossing indicator based on the vertex-wise confidence intervals derived in the first stage to construct the final Hoeffding confidence band.

To construct the mean uncertainty bounds for a confidence level $\beta \in (0,1)$, we want the probability of error to be at most \(1 - \beta\):
\begin{equation}\label{eq:hoeffding-ineq-2}
\Pr\left(\left|\hat{\mu}_n - \mu\right| \ge \epsilon_n\right) \le 1-\beta.
\end{equation}
Using Hoeffding's inequality in \cref{eq:hoeffding-ineq} and \cref{eq:hoeffding-ineq-2}, we set
\begin{equation}\label{eq:hoeffding-ineq-3}
2\exp\left( -\frac{2n\epsilon_n^2}{(b-a)^2} \right) = 1-\beta.
\end{equation}
Solving for \(\epsilon_n\) in \cref{eq:hoeffding-ineq-3}, we obtain
\begin{equation}\label{eq:epsilon_n}
\epsilon_n(\beta) = (b-a) \sqrt{\frac{\ln(2/(1-\beta))}{2n}}.
\end{equation}
The quantity \(\epsilon_n(\beta)\) in \cref{eq:epsilon_n} is the variation away from the sample mean, such that with probability at least \(\beta\), the (usually unknown) ground truth mean \(\mu\) lies in the mean uncertainty interval
\begin{equation}\label{eq:uncertainty-range}
\mu \in \left[ \hat{\mu}_n - \epsilon_n(\beta), \hat{\mu}_n + \epsilon_n(\beta) \right]  = \left[ L , U \right].
\end{equation}
\rev{When $a$ and $b$ are not provided by a priori assumptions, we estimate them from the local sample minimum and maximum values.}
The second stage identifies cells that may contain the isocontour under this uncertainty model. For a given cell \(C_j\), let \([L_k,U_k]\) denote the uncertainty interval associated with the corresponding vertices \(v_k\) of the cell. For a prescribed isovalue \(\tau\), the cell-crossing indicator \(\rho_j\) is defined as:
\begin{equation}
\label{eq:cell-crossing}
\rho_j =
\begin{cases}
1, & \tau \in \bigcup\limits_k [L_k,U_k], \\
0, & \text{otherwise}
\end{cases}
\end{equation}
\rev{where $j$ denotes the cell index and $k \in \{0,1,2,3\}$ indexes the corresponding vertex $v_k$ of each cell.}
\rev{Thus, a cell is included when the isovalue $\tau$ lies within the uncertainty interval of any of its vertices, and the union of these cells forms the confidence band.}

\begin{figure}[htb!]
    \centering
    \includegraphics[width=0.9\linewidth]{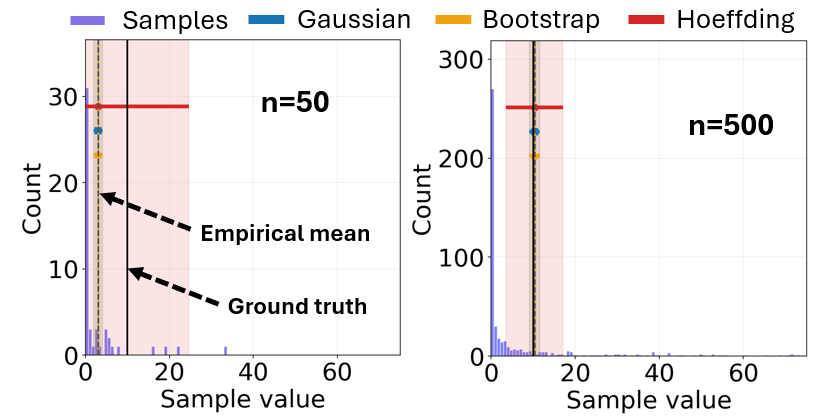}
    \vspace{-5mm}
    \caption{
    Robust Hoeffding confidence interval for truncated skewed/log-normal distribution with $n=50$ (left) and $n=500$ (right) samples. For $n=50$, the Gaussian (blue) and bootstrap (orange) confidence intervals are smaller and do not include the ground truth (vertical black line), thereby providing misleading uncertainty representation. In contrast, Hoeffding bounds (red) encompass the ground truth. As the sample count increases to $n=500$, the bounds for all three methods get compact and encompass the ground truth.
    }
    \label{fig:bimodal-skewed}
\end{figure}


\Cref{fig:bimodal-skewed} validates our proof (\cref{eq:uncertainty-range}) of confidence intervals and showcases how their theoretical robustness bounds the truth. In \cref{fig:bimodal-skewed}, samples are drawn from a truncated skewed/log-normal distribution with bounds $[a,b]=[0,150]$ and ground truth \(\mu = \mathbb{E}[X]= exp(0.5\times4^2) \frac{\Phi(\log(b)/4 -4)-\Phi(\log(a)/4 -4)}{\Phi(\log(b)/4)-\Phi(\log(a)/4)}\). $\Phi$ is the standard Gaussian cumulative distribution function (CDF). The truncated range ensures that the distributions are bounded and the data are generated using sampling with rejection of values outside the bounds. 

In \cref{fig:bimodal-skewed}, for $n=50$, (left image), the Hoeffding confidence interval (red) is wider than the Gaussian (blue) and empirical bootstrap (orange) confidence intervals, but reliably bounds the ground truth. Specifically, as it can be observed in \cref{fig:bimodal-skewed} for $n=50$, the ground truth $\mu$ (vertical black line) lies within the Hoeffding confidence interval, whereas the Gaussian and bootstrap confidence intervals do not overlap with the truth and hence are misleading for a small sample count. Thus, the theoretical nature of the Hoeffding confidence interval makes it a robustness baseline, especially when the sample count is small in critical science applications. As the sample count grows to $n=500$ (right image in~\cref{fig:bimodal-skewed}), the Hoeffding confidence interval becomes more compact. The confidence intervals generated by the Gaussian (blue) and bootstrap (orange) models are subsets of the Hoeffding (red) confidence interval, and all three bound the ground truth $\mu$.

\section{Results}
\label{sec:results}
We evaluate our method on three examples: a synthetic circular level set, a wind magnitude ensemble, and a K\'{a}rm\'{a}n vortex ensemble. In each case, we compare against Gaussian and bootstrap confidence bands using the cell-crossing criterion derived in \cref{eq:cell-crossing}.



\subsection{Synthetic Circle Example}
\label{subsec:circle-results}

\begin{figure}[htb!]
    \centering
    \includegraphics[width=0.88\linewidth]{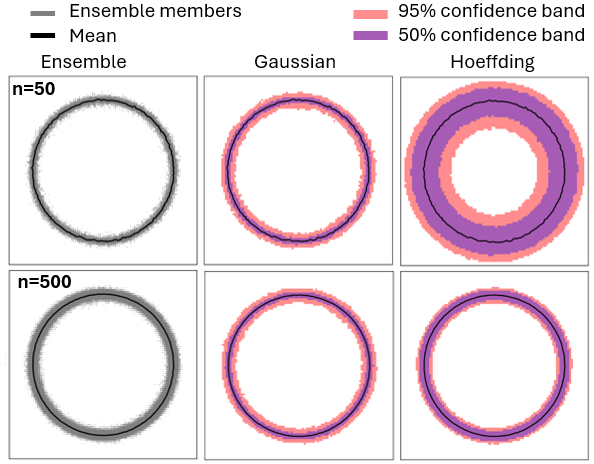}
    \vspace{-4mm}
    \caption{
    Confidence bands comparison for sample counts $n=50$ (top row) and $n=500$ (bottom row). For $n=50$, the Hoeffding band is substantially looser compared to the Gaussian bands. As the sample count increases to $n=500$, the confidence bands become narrower and more localized around the mean isocontour.
    }
    \label{fig:circle}
\end{figure}

The circle ($f_i(x,y) = x^2 + y^2 - 0.75^2 + \eta_i(x,y)$) provides a controlled setting in which the truth contour is known.
The noise $\eta_i(x,y)$ is modeled as a truncated Gaussian distribution with the zero mean, variance $0.0025$, and truncation bounds $[-0.25, 0.25]$. \rev{This provides a controlled setting in which $b-a = 0.5$ is fixed.} \Cref{fig:circle} compares the ensemble contours, mean contour, and confidence intervals obtained using the Gaussian and Hoeffding models for $n=50$ (top row) and $n=500$ (bottom row) samples. 

The mean contour (black) provides a deterministic summary, but it does not indicate spatial uncertainty. For $n=50$, the Hoeffding confidence band (rightmost image) produces a loose envelope compared to the Gaussian confidence band (center image). The Hoeffding confidence band width decreases and closely matches with the Gaussian confidence interval width, as the number of samples increases from $n=50$ to $n=500$. \rev{The Hoeffding confidence band can exceed the empirical ensemble boundary because it is computed from the range $[a,b]$, rather than the observed sample spread. Thus, for $n=50$, the bound is loose; its width decreases as $n^{-1/2}$ and becomes tighter as the ensemble size increases.}

In this controlled example, the Gaussian confidence bands are compact and near the mean contour, which is the expected result given that the assumptions are appropriate for the presented synthetic case. The Hoeffding band, however, serves as a loose but reliable theoretical reference envelope, indicating areas of extra caution where the Gaussian and Hoeffding confidence bands do not agree, especially for a smaller sample count (the top row in \cref{fig:circle}). 

\begin{figure*}[htb!]
    \centering
    \includegraphics[width=0.90\linewidth]{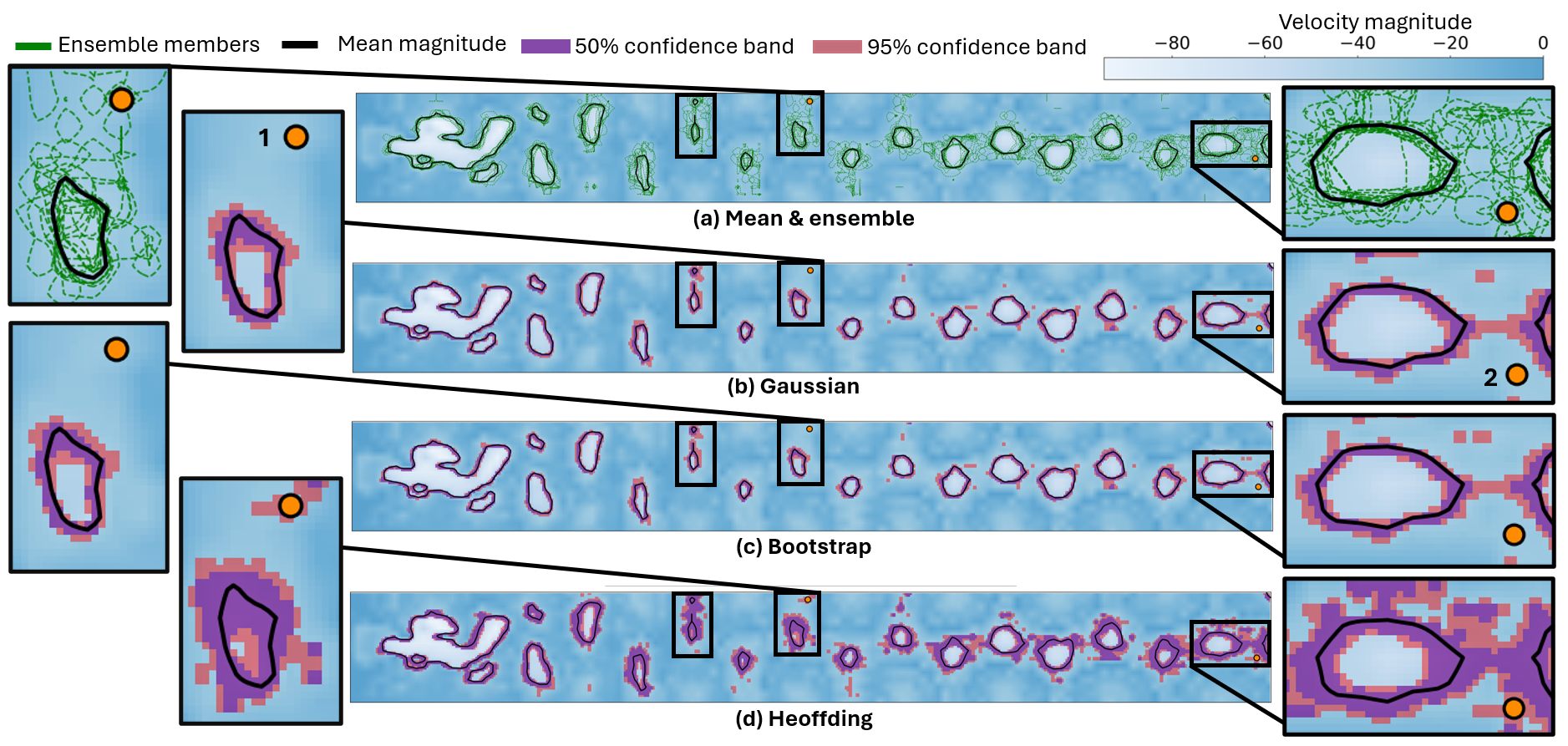}
    \vspace{-5mm}
    \caption{Demonstration of enhanced confidence-band representation with the proposed Hoeffding method for the K\'{a}rm\'{a}n vortex ensemble dataset. Row (a) shows the mean flow velocity magnitude colormapped in blue with ensemble contours (dotted green) and the mean isocontour (black). 
    The Gaussian (row (b)) and bootstrap (row (c)) confidence bands are compact and closely follow the mean isocontour, but suppress spatial uncertainty in regions where ensemble contours exhibit scattered behavior. In contrast, our derived distribution-agnostic Hoeffding confidence bands (row (d)) are wider in multiple regions and robustly capture ensemble contour variability (green contours) seen in row (a).}
    \label{fig:karman-vortex}
\end{figure*}

\subsection{Wind Dataset}
\label{subsec:wind-results}
We evaluate the proposed Hoeffding confidence band method on a real ensemble of wind velocity fields comprising $15$ members~\cite{Vitart2017}. We compute the wind magnitude at each grid vertex for each ensemble member. The small sample size ($n = 15$) and the absence of any known generating distribution are exactly the conditions under which the Hoeffding confidence band is intended to be most useful and robust relative to the Gaussian or bootstrap alternatives.

\Cref{fig:teaser} shows a comparison of confidence band visualizations at the isovalue $10.0$. \Cref{fig:teaser}(a) shows the mean (black) and ensemble (green) isocontours. The results in \cref{fig:teaser}(b)-(d) show the confidence bands from the Gaussian, bootstrap, and proposed Hoeffding models, respectively. \Cref{fig:teaser}(f) shows zoomed-in views of a selected local region. \Cref{fig:teaser}(e) shows a histogram plot of the $15$ ensemble members at a selected point (orange dot). 
%


The $50\%$ and $95\%$ confidence bands visualized in \cref{fig:teaser}(b)-(d) show how the spatial uncertainty region expands as the required confidence level increases. As expected, the confidence bands from the Gaussian and bootstrap methods (depicted in orange and blue, respectively) are significantly more compact than the Hoeffding-based method (depicted in red). As observed from the zoomed-in views in ~\cref{fig:teaser}(f), these tighter bands fail to capture contour structures that are visible in the ensemble isocontour plot (depicted in green). The Hoeffding confidence interval, although loose, better bounds the green ensemble isocontour structures and theoretically indicates that the unknown ground truth lies in these bands. 

The difference between the Hoeffding and distribution-driven/empirical bands is particularly important in this low sample count ($n=15$) case where estimate of the true distribution shape is difficult to capture, as previously demonstrated in \cref{fig:bimodal-skewed}. The histogram in \cref{fig:teaser}(e) indicates a case where the tighter Gaussian and bootstrap intervals do not bound the isovalue, whereas the Hoeffding interval remains wide enough due to its theoretical nature to bound it. The proposed Hoeffding method is computationally efficient ($20$ ms) with a runtime comparable to Gaussian ($21$ ms) and faster than bootstrap ($78$ ms).

\subsection{K\'{a}rm\'{a}n Vortex Dataset}
\label{subsec:karman-results}

The K\'{a}rm\'{a}n vortex street ensemble generated with the Gerris software~\cite{POPINET2003572} consists of $15$ flow simulations with varying viscosity parameter. Unlike the synthetic and wind field examples, which contain few closed distinct contours, K\'{a}rm\'{a}n vortex consist of many small, spatially separated contours due to its turbulent nature, thus exhibiting significant variations among the ensemble members. 

\Cref{fig:karman-vortex} shows results similar to \cref{fig:teaser}. Specifically, it compares the mean and ensemble contours (\cref{fig:karman-vortex}(a)), and confidence bands obtained from the Gaussian (\cref{fig:karman-vortex}(b)), bootstrap (\cref{fig:karman-vortex}(c)), and Hoeffding (\cref{fig:karman-vortex}(d)) methods at the isovalue $-40.0$. The ensemble contours (dotted green) reveal substantial contour position variability across the wake. Directly overlaying all ensemble contours produces a cluttered spaghetti plot that makes it difficult to assess the confidence of the mean isocontour (black).


\begin{figure}[htb!]
    \centering
    \includegraphics[width=0.6\linewidth]{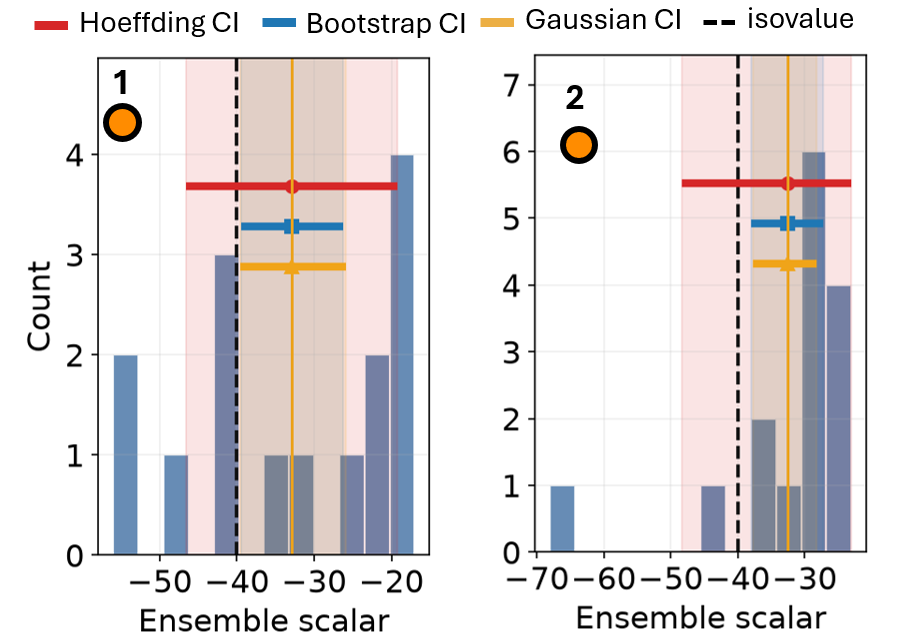}
    \vspace{-3mm}
    \caption{
    Diagnostics of K\'{a}rm\'{a}n  vortex street ensemble. The orange dots (annotated as $1$ and $2$) mark the selected locations in \cref{fig:karman-vortex}, and the intervals show $95\%$ confidence interval. The black dashed line indicates the chosen isovalue. The Hoeffding interval (red) is wide enough to bound the isovalue, demonstrating how the Hoeffding band captures possible contour crossings missed by the more compact Gaussian (orange) and bootstrap (blue) intervals.
    }
    \label{fig:karman-vortex-histogram}
\end{figure}

As seen from \cref{fig:karman-vortex}, the Gaussian and bootstrap confidence bands compactly enclose the mean isocontour at each individual vortex structure, creating disjoint bands on the right side of the spatial domain. Thus, such compactness suppresses and underestimates spatial uncertainty. In contrast, the proposed Hoeffding confidence band method produces wider confidence regions around vortical structures, where these bands begin to merge, especially on the right side of the flow domain, and exhibit a consistent merging behavior of the green ensemble isocontours seen in \cref{fig:karman-vortex}(a). This behavior is a direct consequence of the theoretical and distribution-agnostic nature of the Hoeffding confidence band. It can be viewed as a highly desirable safety property where vortex structures are closely spaced and sample sizes are limited, especially in critical applications such as vortex formation in tall urban structures~\cite{irwin2010vortices}.



In \cref{fig:karman-vortex-histogram}, the two orange dots mark the selected grid vertices in \cref{fig:karman-vortex}, and the histograms correspond to 15 ensemble values at these vertices, overlaid with the $95\%$ confidence intervals from the Hoeffding, Gaussian, and bootstrap models. The black dashed line indicates the isovalue. At both locations, the histograms suggest that the isovalue lies within a plausible uncertainty range. The proposed, theoretical Hoeffding confidence interval reliably bounds the isovalue, whereas it is missed by the more compact Gaussian and bootstrap confidence intervals.

\section{Conclusion and Future Work}
We introduce a Hoeffding confidence band technique that provides a theoretical, distribution-agnostic isocontour uncertainty visualization. Experiments on synthetic and real ensemble datasets show that the method produces, although less localized, more robust confidence bands than the Gaussian and empirical bootstrap alternatives, reliably capturing uncertain contour regions with a theoretical guarantee of bounding the unknown truth. The Hoeffding band computation remains efficient with runtime comparable to Gaussian and faster bootstrap. Rather than competing with existing methods, this approach complements them by providing a robust, distribution-agnostic baseline for small ensembles and safety-critical applications where reliable uncertainty estimates are desirable. In terms of limitation, the uncertainty bands may become too wide for the Hoeffding method. Future work will explore tighter theoretical, and distribution-agnostic confidence bounds. We will also explore extensions to 3D volumetric datasets and their topological features beyond isocontours, such as critical points and Morse complexes.
%
\acknowledgments{
This work was supported by the U.S. Department of Energy (DOE), Office of Science, Office of Advanced Scientific Computing Research (ASCR), Early Career Research Program under Contract DE-AC0500OR22725. The authors also acknowledge research sponsored by the Laboratory Directed Research and Development (LDRD) Program of Oak Ridge National Laboratory, managed by UT-Battelle, LLC, for the U. S. DOE.}
\bibliographystyle{abbrv-doi}

\bibliography{references}
\end{document}